\documentclass[prl,bibnotes,twocolumn]{revtex4}
\usepackage{graphicx}
\begin{document}
\draft
\def\ds{\displaystyle}
\title{ Spontaneous topological pumping in non-Hermitian systems }
\author{C. Yuce}
\address{Department of Physics, Eskisehir Technical University, Turkey }
\email{cyuce@eskisehir.edu.tr}
\date{\today}
\begin{abstract}
We introduce modulational instability in non-Hermitian systems to study state conversion of topological edge states. We show that state conversion in non-Hermitian systems leads to topological pumping, which is a way of transferring topological edge state from one edge to the opposite edge. In contrast to Hermitian systems, topological pumping can occur spontaneously in non-Hermitian systems. 
\end{abstract}
\maketitle

\section{Introduction}

The physics of non-Hermitian systems has seen tremendous growth in the past two decades. Of special importance in non-Hermitian systems is the existence of exceptional points (EPs) \cite{EP1}.  An EP is a phase transition point in parameter space of a non-Hermitian Hamiltonian where at least two eigenvalues and the corresponding eigenvectors of the Hamiltonian coalesce. An interesting behavior around EPs is the state conversion. It was shown that chiral state conversion between eigenstates of a two-level system occurs, regardless of initial states when the EP is dynamically enclosed in parameter space \cite{dynexcep1,dynexcep2}. This implies violation of the adiabatic approximation in non-Hermitian systems.  \\
In recent years, the concept of topological insulating phase has been extended to non-Hermitian systems and this new field of study has attracted great deal of attention. It was shown that topological edge states with real energy eigenvalues can exist in the non-Hermitian Aubry-Andre model \cite{nonh2}. In 2017, topological zero energy edge state in $1D$ lossy waveguides was experimentally realized \cite{sondeney1}. In $1D$, Su-Schrieffer-Heeger (SSH) model with gain and loss is commonly used to explore new physics in the theory of non-Hermitian topological insulators \cite{1d1,1d3,1d3ekl,1d4,1d5,1d6,1d7,1d8,1d9,1d10,1d11,1d12}. Recently, few papers appeared to explore topological phase in $2D$ systems \cite{2d1,2d2,2d3}. It is surprising that standard bulk boundary correspondence does not work in non-Hermitian systems \cite{bulkboun01,bulkboun02,bulkboun03,bulkboun04,bulkboun05,bulkboun06,bulkboun07}. It was shown that the topological phase transition point may not be predicted using periodical form of non-Hermitian Hamiltonians. In other words, systems with open edges have different topological phase transition point than the point predicted by periodical form of the non-Hermitian Hamiltonian. Furtermore, standart formulation of topological numbers such as winding and Chern numbers are not fully compatible in non-Hermitian systems \cite{winding1,winding2,winding3}. Recently, the concept of pseudo topological insulators have been introduced and applied to a non-Hermitian system \cite{yuce383}. Topological systems other then topological insulators have also been studied. These are topological superconductors with gain and loss and Majarona modes \cite{majo1,majo2,majo3,majo4,majo5,majo6,majo7}, nodal surfaces formed by EPs \cite{nodalkun,nodalkun2} and Floquet topological phase in non-Hermitian systems \cite{floquet1,floquet2,floquet3,floquet4}.\\
In this paper, we study state conversion of topological edge states in non-Hermitian systems. We explore modulational instability (MI) to study this phenomenon. We show that state conversion in non-Hermitian systems leads to topological pumping, which is a way of transferring topological edge state from one edge to the opposite edge \cite{pumpy001,pumpy002}. The Rice-Mele model is generally used in Hermitian systems to study topological charge pumping \cite{ricemele,pump001,pump002}. The charges are transported by adiabatic cycling of some parameters in the Rice-Mele model. We show that the topological pumping in non-Hermitian systems can be realized without having to dynamically controlling any parameters of the Hamiltonian. The topological pumping introduced here occurs spontaneously and has no analogue in Hermitian systems. 

\section{Formalism}

We start with a two-level system and then make a generalization. Consider the following two-level time-independent system with gain and loss
\begin{equation}\label{yudj2}
\mathcal{H}=\left(\begin{array}{ccccc}i \gamma_0  & -1  \\ -1  & -i\gamma_0   \end{array}\right)
\end{equation}
where $\ds{\gamma_0}$ is the gain/loss strength. The tunneling amplitude is set to $-1$ for simplicity. Here we study the case with $|\gamma_0|>1$, i. e., spontaneously broken parity-time $\mathcal{PT} $ symmetric region, where the eigenvalues are purely imaginary. The unnormalized eigenstates are given by
\begin{eqnarray}\label{oawipjj2}
\psi_\mp =\left(\begin{array}{cc} - i\gamma_0\pm  \sqrt{1-\gamma_0^2}  \\ \ 1   \end{array}\right)  
\end{eqnarray}
and the corresponding eigenvalues read $\ds{ E_\mp=\mp i E_I}$, where $\ds{E_I=\sqrt{\gamma_0^2-1} }$. Note that $\psi_+$ grows in time while $\psi_-$ decays. Of special importance is the EPs at $\gamma_0={\mp}1$, where $\psi_+=\psi_-$ and $E_+=E_-=0$. \\ 
Our aim is to study the stability of these two eigenstates. Let us now add perturbation (amplitude modulation) to them. Suppose that the modulated eigenstates are given by $\ds{\psi_{\mp}^{\prime}=\psi_\mp+\delta\psi  }   $, where $\ds{\delta\psi=\left(\begin{array}{cc}   \epsilon_1 \\ \  \epsilon_2  \end{array}\right)  }   $ and $\ds{\epsilon_{1,2}}$ are very small complex numbers, $\ds{|\epsilon_{1,2}|<<1}$. Note that such a weak perturbation always exists in a real physical system as it is not possible to eliminate noise in an experiment. Besides, even very small tunneling amplitude disorder in an experiment is equivalent to such a weak perturbation. A question arises. Are the two eigenstates (\ref{oawipjj2}) stable against the amplitude modulation? If they are not, we can say that the system is modulationally unstable. MI which leads to exponential growth of the weakly perturbed wave is one of the most fundamental effects in nonlinear systems \cite{NLSkonotop}. However, MI studied here is not exactly the same as the one in nonlinear system. Fortunately they share some common features such as exponential growth of the modulation. Below we will show that $\ds{\psi_+}$ is modulationally stable while $\ds{\psi_-}$ is not. \\
To explore the dynamics in our system, we will find the time evolution of an arbitrary initial state $\ds{|\psi(t=0)>=\{a_0,b_0\}^{T}}$, where $\ds{a_0}$ and $\ds{b_0}$ are complex numbers. We are mainly interested in the dynamics at large times. Let us study time evolution at large times in two cases: i-) $\ds{\gamma_0={\mp}1}$ and ii-) $\ds{|\gamma_0|>1}$. In the first case, the state at large times is always the exceptional state, $\ds{|\psi(t)|>=\{1,{\mp}i\}^{T}}$, regardless of the initial state. This is because there is only one eigenstate at the EP, and any initial state will eventually be the exceptional state. In the second case, the eigenvalues are purely imaginary. Therefore, the system prefers the eigenstate with the higher eigenvalue. Too see this, we solve the equation $\ds{\mathcal{H}|\psi(t)>=i\partial_t |\psi(t)>}$ analytically \cite{cozum}. We get the time dependent state vector at large times, which is insensitive to the initial conditions
\begin{eqnarray}\label{o987jj2}
\psi(t>>0)\sim\frac{  ia_0 + (E_I  -  \gamma_0) b_0   }{2 E_I}   ~   \exp{( E_I t)} ~ \psi_+
\end{eqnarray}
It is interesting to see that the initial values $a_0$ and $b_0$ become just a multiplicative factor at large times. This means that the state is always in the state $\psi_+$, regardless of the initial state. The solution (\ref{o987jj2}) is valid unless $\ds{ ia_0 + (E_I  -  \gamma_0) b_0=0}$ and $E_I=0$. The former one is satisfied if the initial state is $\psi_-$ as can be seen from (\ref{oawipjj2}) and the latter one occurs at the EP. In the former case, the time-dependent state vector is given by $\ds{\psi(t)=  \exp{( -E_I t)} ~ \psi_-}$. Let us consider the modulation of $\psi_-$: $\ds{\psi_{-}^{\prime}=\psi_-+\delta\psi  }   $.  Then MI leads to state conversion and the state vector at large times is now given by (\ref{o987jj2}). The time required for the state conversion depends on the $\delta\psi$ and $E_I$. Note that the modulated wave $\ds{\psi_{-}^{\prime}}$ grows in time (at large times) although $\ds{\psi_{-} }$ decays in time. We stress that no such effect can be observed for the modulated $\ds{\psi_+}$. We conclude that the state tends to follow the eigenstate with less loss. Thus only one eigenstate effectively survives at large times. Note that this leads to the nonreciprocity effect \cite{nonrecip}.\\
Let us now consider a two state system with unbalanced gain and loss. The Hamiltonian can then be given by
$\ds{\mathcal{H}^{\prime }=-i\gamma_1\mathcal{I}+\mathcal{H}}$, where $\ds{\mathcal{I}}$ is the identity matrix and $\ds{\mathcal{H}}$ is described in (\ref{yudj2}) and $\gamma_0>1$, $\gamma_1$ is a free parameter. Consider first $\gamma_1=\gamma_0$ for which the second site has loss. Since the Hamiltonian is just shifted, the time-dependent state at large times is given by $\ds{\psi^{\prime}(t>>0)=\exp{(-{\gamma_0}t)}\psi(t>>0)}$. Therefore the corresponding density decays in time according to $\ds{ |\psi^{\prime}(t>>0)|^2\sim    \exp{(2(-{\gamma_0}+E_I)t)}}$. One can see that $-\gamma_0+E_I$ decreases with $\gamma_0$ and is practically zero for very large values of $\gamma_0$. This is interesting since the decaying rate decreases with increasing loss. For two coupled waveguides, this is equivalent to the enhancement of the transmission as loss is increased \cite{nonrecip}. Let us consider another special case, $\gamma_1=E_I$ for which both sites have losses. In this case the wave-packet neither grows nor decays for any $\gamma_0>1$ at large times.\\
In Hermitian systems, if the initial states is one of the eigenstates, then it remains in that stationary eigenstate. In an experiment with the Hermitian two-level system, unavoidable noises have only perturbational effects. However, this is not the case in non-Hermitian systems as MI plays a vital role. We remark that this effect can be seen not only in the above $2$-level system but also in other non-Hermitian systems. The modulationally stable state in an $N$-level non-Hermitian system is the eigenstate with the eigenvalue whose imaginary component is the highest in the system. All other states are modulationally unstable and evolve to that state. If there are more than one such eigenstates (multiple degeneracy in the highest imaginary part of eigenvalues), then the state at large times is a linear combination of these states.\\
Below, we apply our idea to a non-Hermitian topological system and discuss topological state conversion in a non-Hermitian lattice. We will show that a novel topological pumping effect occurs. We stress that the topological pumping in our system occurs spontaneously (it happens without dynamical change of system parameters). This has no anoloque in Hermitian systems. 

\section{Topological state conversion}

Consider the well known 1D SSH tight-binding chain \cite{ssh}. We add either gain or loss to the chain. Suppose gain/loss are absent on odd values of lattice sites while they present on even values of lattice sites. The periodical form of the total Hamiltonian is then given by
\begin{equation}\label{mwedfaz6}
\mathcal{H}=\left(\nu+\omega\cos(k) \right)\sigma_x+\omega\sin(k)\sigma_y+i\frac{\gamma_0}{2}(\mathcal{I}-\sigma_z)
\end{equation}
where $\ds{\mathcal{I}}$ is the identity matrix, $\vec{\sigma}$ are Pauli matrices, the crystal momentum $k$ runs over the first Brillouin zone, $-\pi<k<\pi$ and the real-valued positive parameters $\ds{\nu>0}$, $\ds{\omega>0}$ are hopping amplitudes. The system has only gain (loss) if $\gamma_0>0$ $(\gamma_0<0)$. Such a system has topological zero energy state \cite{sondeney1}. \\
The corresponding energy eigenvalues are given by $\ds{E_{\mp}=i\gamma_0/2\mp\sqrt{ \nu^2+\omega^2+2~\nu~\omega  \cos(k) -(\gamma_0/2)^2 }}$, which is symmetrically located around the energy $i\gamma_0/2$. As it is well-known, bulk-boundary correspondence doesn't work in non-Hermitian systems. Therefore the periodical system and finite system with open edges have different spectra and topological phase transition points. Let us study the finite system with open edges. Suppose that the number of lattice sites $\ds{N}$ is an even number. Therefore gain/loss is absent at the left edge ($n=1$ lattice site) while it presents at the right edge ($n=N$ lattice site). Consequently, the parity symmetry is lost in the system and hence the two edge states are no longer symmetric. We expect that the left edge state has exactly zero eigenvalue while the right edge state has a purely imaginary eigenvalue whose sign is positive (negative) when $\ds{\gamma_0>0}$ $\ds{(\gamma_0<0)}$. Therefore, the topological left (right) edge state is certainly modulationally unstable if the system has gain (loss). To determine whether the topological edge state at the other edge is modulationally stable or not, we should also find the eigenvalues of the bulk states. If the imaginary part of its energy eigenvalue is highest in the system, then all other eigenstates are converted into the topological edge state at large times. More generally, final state at large times is the topological state localized around one edge, regardless of the initial state and its topological character. Thouless introduced the idea of topological pump in which charges are transported by adiabatic changing of a time-periodic Hermitian Hamiltonian \cite{pump000}. In our system, an analog of topological pumping occurs due to MI without changing some parameters of the Hamiltonian. \\
To confirm our predictions, we perform numerical simulations for the parameters $\ds{N=20}$, $\ds{\gamma_0=0.2}$, $\ds{\omega=0.6}$ and $\ds{\nu=1.4}$. We numerically find that the left and right edge states have the eigenvalues $E=0$ and $E=2i\gamma_0$, respectively. The real part of the bulk eigenstates come in pairs with $\ds{{\mp}E_R}$ and their imaginary parts are all equal to $\ds{E_I=i\gamma}$. Let us numerically find the eigenstates and add very small noises. Then we study their time evolutions. In Fig.1, we plot the normalized densities $\ds{|c_n|^2/P}$, where $c_n$ is the complex amplitude at the site $n$ and $P=\sum_n |c_n|^2$ is the total intensity. In Fig.1 (a), the initial state is the weakly modulated left edge state. As can be seen from the figure, state conversion occurs at around $\ds{t=20}$, which decreases with increasing $\gamma_0$ and reducing the system-size. Once the state conversion is initiated at around $t=20$, the conversion of the left edge state into the other edge state occurs very fast. This spontaneous topological pumping is novel and impossible to observe in Hermitian systems, where edge state transport can only be possible if some parameters of the Hamiltonian such as tunneling amplitude and staggered potential are varied in time \cite{ricemele}. Consider next that the initial state is the weakly modulated right edge state. As can be seen from the Fig. 1.(b), this state is modulationally stable and propagates without being converted into any other eigenstates. Furthermore, it is topological and resists to some kind of disorders as will be studied below. Let us now study time evolution of bulk states. In Fig.1 (c,d), one can see how two bulk states are transported to the right edge and finally converted into the topological right edge state. These are in total agreement with our predictions. To this end, we note that similar behavior can still be observed under replacement $\ds{\gamma\rightarrow-\gamma}$. In this case, any wave packet is transported to the left edge. The total power decreases during transportation and eventually stops decaying once it is converted to the topological edge state. 
\begin{figure}[t]
\includegraphics[width=4.25cm]{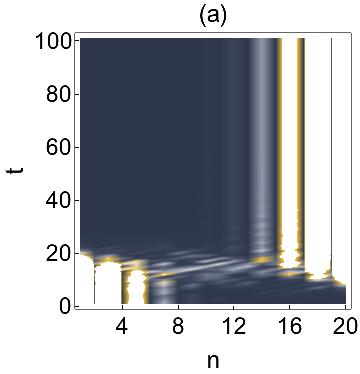}
\includegraphics[width=4.25cm]{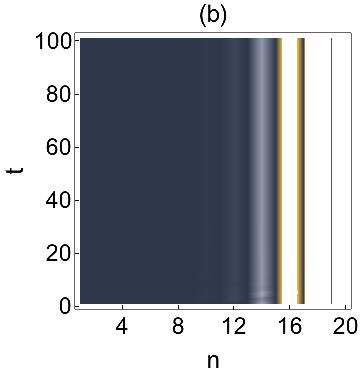}
\includegraphics[width=4.25cm]{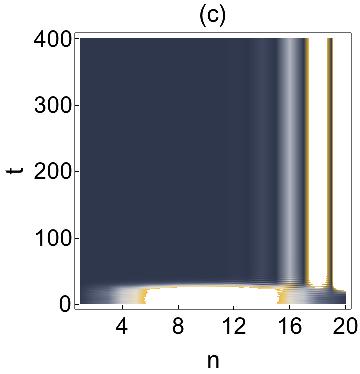}
\includegraphics[width=4.25cm]{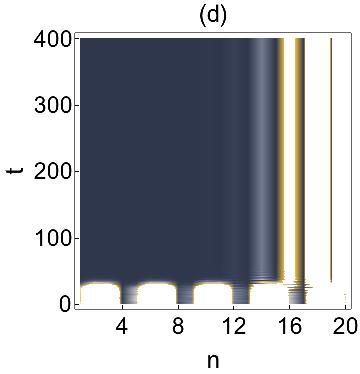}
\caption{The (normalized) density plots as functions of $n$ and time at $\ds{\gamma=0.2}$ for topological left and right edge states in (a,b) and for bulk states with eigenvalues $E=1.97 + i\gamma$ and $E=1.53 + i\gamma$ in (c) and (d), respectively.  We initially add very small amplitude modulation to these states. The topological right edge state is the only modulationally stable state in the system. It is interesting to see that topological left edge state is transported to the other edge and all the bulk states evolve to the topological state at the right edge. The state conversion occurs at around $t=20$, which decreases rapidly with increasing $E_I$. In (c,d), we plot the figures up to $t=400$ to confirm that no other state conversion occurs.}
\end{figure}\\
Robustness of topological edge states against certain types of disorder is of great importance in the theory of topological insulating systems. Let us now study robustness of the topological edge states in our system against two types of disorders. These are tunneling amplitude disorder and gain/loss strength disorder. Consider first the former case. We introduce randomized coupling all over the lattice such that $\ds{\omega\rightarrow\omega+\delta\omega_n}$ and $\ds{\nu\rightarrow\nu+\delta\nu_n}$, where $\ds{\delta\omega_n<<\omega}$  and $\ds{\delta\nu_n<<\nu}$ are real-valued random set of site dependent tunneling amplitudes in the interval $\ds{[-0.1,0.1]}$. Consequently, the tunneling amplitudes between two neighbouring sites become completely independent. We numerically see that the real parts of the energy eigenvalues for the right and left edge states remain the same in the presence of the disorder while they are sensitive to the disorder for the bulk states. Note that the left edge state is still topological but modulationally unstable. The imaginary part of eigenvalues don't change with the disorder for all states. Therefore, the right edge state has the highest imaginary part of eigenvalue and modulationally stable. Consider next the gain/loss strength disorder. In this case, we introduce extra randomized gain and loss such that $\ds{\gamma_0\rightarrow \gamma_0+\delta\gamma_{n}}$, where $\ds{    \delta \gamma_n  \in  [-\gamma_1 , \gamma_1]}$. The right edge state has still zero energy since it resists to the disorder while only the real part of energy eigenvalue for the left edge state resists to the disorder. However, both the real and imaginary parts of energy eigenvalues for the bulk states change considerably with the disorder. Fortunately, the right edge state has always the highest imaginary part of energy eigenvalue as long as $\gamma_0>\gamma_1$. Therefore, the right edge state is still the modulationally stable state.  \\
An intuitive explanation for the spontaneous topological pumping is as follows. Suppose imaginary parts of energy eigenvalues of all eigenstates are negative. Then any initial state decays in time but a small amount of energy leaks to the state with highest imaginary part of energy eigenvalues, which eventually dominates and decays more slowly as the loss is increased. A similar discussion can be made if the system has gain. In this case the state with more gain grows more rapidly relative to the other states and becomes dominant at large times.\\
To sum up, we study state conversion of topological edge states in non-Hermitian systems and show that final state at large times is the topological state localized around one edge, regardless of the initial state and its topological character. In contrast to transport in Hermitian systems, the topological pump in non-Hermitian system can be spontaneous. Topological pumping in Hermitian systems usually takes a long time since it is adiabatic \cite{pump001}. In non-Hermitian system, this occurs faster. The topological pumping introduced here is due to the MI and has no analogue in linear Hermitian systems. This effect can be used for designing a topological laser as the output wave at one edge is always single mode even if the input wave is an extended multimode wave.


\begin{thebibliography}{0}
\bibitem{cozum} The exact solution is given by $\ds{|\psi(t)>=\{a(t),b(t)\}^{T}}$ where $\ds{  a(t)= a_0 \cosh{( E_I  t)}  +\frac{  \gamma_0 a_0 +ib_0  }{E_I  } \sinh{(E_I  t )}  }$ and $\ds{
b(t)=b_0 \cosh{(E_I t)}  +i \frac{  a_0+ i \gamma_0 b_0    }{E_I  } \sinh{(E_I  t)} }$
\end{thebibliography}
\end{document}